\documentstyle[aps,prl,preprint]{revtex}

\title{Statistical Inference, Distinguishability of Quantum States, And 
Quantum Entanglement}
\author{V. Vedral, M.B. Plenio, K. Jacobs and P. L. Knight}
\address{Optics Section, Blackett Laboratory, Imperial College London, 
London SW7 2BZ, England}
\date{\today}

\begin{document}

\maketitle

\begin{abstract}
We argue from the point of view of statistical inference that the quantum 
relative entropy is a good measure for distinguishing between two quantum 
states (or two classes of quantum states) described by density matrices.
We extend this notion to describe the amount of entanglement between two 
quantum systems from a statistical point of view.  Our measure is independent 
of the number of entangled systems and their dimensionality. 
 
\end{abstract}

Recent work has taught us that Bell's inequalities are not always a
good criteria for distinguishing entangled states (i.e. those possessing a 
degree of quantum correlations) from 
disentangled states \cite{Gisin}. This discovery has initiated much work
in quantum information theory (e.g \cite{Deutsch,Horodecki1}) particularly concerning the search for a measure of the amount of entanglement contained within a given quantum state \cite{Bennett1,Vedral1,Vedral2}. In a recent letter \cite{Vedral2}, we presented conditions that any measure of entanglement has to satisfy. 
This was motivated by the fact that local actions, combined only with classical communications, should not be able to increase the amount of entanglement \cite{Bennett1,Vedral1,Vedral2}. In \cite{Vedral2} we defined our measure 
as the minimal distance of an entangled state to the set of disentangled states.
This distance function (not necessarily a metric) could, for example, be satisfied by the quantum relative entropy (to be defined later) and by the Bures metric (for definition see e.g. \cite{Fuchs}). Our measure of entanglement was derived 
from the abstract idea of closest approximation rather than from intuitive physical grounds. In this paper we start from an entirely different point of view and derive a measure of entanglement from the idea of distinguishing two quantum states starting from classical information theory \cite{Cover}. We find that these new insights lead to the same measure of entanglement as in \cite{Vedral2}
(but now the quantum relative entropy is picked out from among the possible measures of ``distance"). This corroborates the 
results of \cite{Vedral2} and puts them on a firm statistical basis
allowing experimental tests to determine the amount of entanglement.

In order to understand our argument in the quantum case we must first 
describe its classical counterpart. Suppose that we are asked to distinguish
between two probability distributions, taken for simplicity to be discrete. 
Say that
we either have a fair coin with a ``fifty--fifty" head--tail probability distribution, or an unfair coin with ``seventy--thirty" head--tail probability distribution. We are allowed to toss a {\em single} coin $N$ times and we want to know which one it is. To be more general, let us say that we have a dichotomic variable
with the following distribution of probabilities: $p(1)=p$ and $p(0)=1-p$. The
probability that from $N$ experiments (trials) we obtain $n$ 1's and $(N-n)$ 0's is given by the binomial distribution:

\begin{equation}
P_N(n) = {N \choose n} p^n (1-p)^{N-n} \;\; .
\end{equation}
This can be written as 
\begin{equation}
P_N(n) = \exp \{ \ln P_N(n) \} = \exp \{\ln {N \choose n} p^n (1-p)^{N-n} \} \;\; .
\end{equation}   
However using the Stirling's approximation for large numbers the exponent can be considerably simplified:

\begin{eqnarray}
\ln {N \choose n} p^n (1-p)^{N-n} & = & -N \{ \frac{n}{N}\ln\frac{n}{N} + (1-\frac{n}{N})\ln(1-\frac{n}{N}) \nonumber \\
& + & \frac{n}{N} \ln p + (1-\frac{n}{N})\ln(1-p) \}
\label{3}
\end{eqnarray}
Now the quantity $n/N$ is our {\bf measured} frequency of 1's and likewise 
$1-n/N$ is the {\bf measured} frequency of 0's in $N$ trials. The probabilities which we infer from this distribution are given by the maximum likelihood estimate \cite{Cover}: $p_{inf}(1)=n/N$ and
$p_{inf}(0)=1-n/N$ . These are, in general, different to $p$ and $1-p$.
The crucial question we wish to ask, therefore, is: what is the
probability that after $N$ trials our inferred probabilities are $q$ and $1-q$, if the experiment was done using a system having ``true" probabilities $p$ and $1-p$? In the light of the coin example we ask what is the probability of wrongly inferring that we have a fair coin when, in fact, the ``seventy--thirty" unfair one was used in the experiments?  Clearly the answer is given by replacing $n/N$ by $q$ in eq. (\ref{3}). The result in the large $N$ limit is

\begin{equation}
P_N(p \rightarrow q) = e^{-NS(q||p)} \;\; ,
\end{equation}
where
\begin{eqnarray} 
S(q||p)  :=  \{ q\ln q + (1-q)\ln(1-q) 
 -  q \ln p - (1-q)\ln(1-p) \}
\end{eqnarray}
is the so called relative entropy, or the Kullback--Leibler distance \cite{Vedral1,Vedral2,Cover,Wehrl1} between the binary distributions $p$ and $q$. In general
it is easy to see that the probability to confuse a distribution $\{p\}_1^M$
with $\{q\}_1^M$ in $N$ measurements is given by
\begin{equation}
P_N(p \rightarrow q) = e^{-N\sum_i q_i \ln q_i - q_i \ln p_i} \;\; .
\label{conf}
\end{equation}
As the relative entropy is an asymmetric quantity a natural question to ask is: why is
the probability of confusing $p$ with $q$ different to the probability to 
confuse $q$ with $p$? The following simple ``coin" example will explain
this. Suppose we have a fair coin and a completely unfair coin (two--heads
for example). Suppose we have to decide which one it is, but we are allowed to
do $N$ experiments on only {\bf one}, of course unknown--to--us, coin. So,
say we are tossing the unfair coin. Then as heads is the only possible outcome,  we will never confuse the unfair coin with the fair one, as after each trial
the inferred probabilities will be $p(\mbox{head})=1$ and $p(\mbox{tail})=0$.
This is in fact corroborated by our formula in eq. (\ref{conf}) as $e^{-\infty} =0$. On the other hand, suppose we are tossing the fair coin: then after the
first outcome which could equally be heads or tails we have probability of 
$1/2$ of confusing the coins (i.e if the heads shows up we will make the wrong
inference, whereas if the tail shows up it will be the right inference). This
also follows from eq. (\ref{conf}) as $e^{-\ln 2}=1/2$ (note that here 
the formula is correct even for $N$ small!). 

The central aim for us in this paper is to generalize this idea to
distinguish (or, equivalently confuse) two quantum states which are
completely described by their density matrices. To that end, suppose we have
two states $\sigma$ and $\rho$. How can we distinguish them? We can chose a
{\bf P}ositive {\bf O}perator {\bf V}alued {\bf M}easure $\sum_{i=1}^M A_i ={\bf1}$ which generates two distributions via 

\begin{eqnarray}
p_i & = & tr A_i \sigma \\
q_i & = & tr A_i \rho \;\; ,
\end{eqnarray}
and use classical reasoning to distinguish these two distributions. 
However, the choice of POVM's is not unique. It is therefore best to choose that POVM
which distinguishes the distributions most, i.e. for which the relative entropy is largest. Thus we arrive at the following quantity

\[
S_1(\sigma ||\rho) := \mbox{sup}_{\mbox{A's}} \{ \sum_{i} tr A_i \sigma \ln tr A_i \sigma - tr A_i \sigma \ln tr A_i \rho \} \;\; ,
\]
where the supremum is taken over all POVM's. 
The above is not the most general measurement that we can make, however. In 
general we have $N$ copies of $\sigma$ and $\rho$ in the state 
\begin{eqnarray}
\sigma^N & = & \underbrace{\sigma\otimes \sigma ... \otimes \sigma}_{\mbox{total of N terms}}\\
\rho^N & = & \underbrace{\rho\otimes \rho ... \otimes \rho}_{\mbox{total of N terms}} 
\end{eqnarray}
We may now apply a POVM $\sum_{i} A_i ={\bf 1}$ acting on $\sigma^N$ and $\rho^N$. Consequently, we define a new type of relative entropy
\begin{eqnarray}
S_{N}(\sigma ||\rho)  :=  \mbox{sup}_{\mbox{A's}} \{ \frac{1}{N} \sum_{i} tr A_i \sigma^N \ln tr A_i \sigma^N  
 -  tr A_i \sigma^N \ln tr A_i \rho^N \}
\label{lim}
\end{eqnarray}
Now it can be shown that \cite{Donald2}
\begin{equation}
S(\sigma ||\rho) \ge S_{N}
\label{ineq}
\end{equation}
where 
\begin{equation}
S(\sigma ||\rho) := tr (\sigma \ln \sigma - \sigma\ln\rho)
\end{equation}
is the quantum relative entropy
\cite{Vedral1,Vedral2,Wehrl1,Donald2,Lindblad1,Lindblad2} (for the summary of the properties of quantum relative entropy see \cite{Ohya}). Equality
is achieved in eq. (\ref{ineq}) iff $\sigma$ and $\rho$ commute \cite{Fuchs1}.
However, for any $\sigma$ and $\rho$ it is true that \cite{Petz} 

\[
S(\sigma ||\rho) = \lim_{N\rightarrow \infty} S_{N} \;\; .
\]
In fact, this limit can be achieved by projective measurements which are
independent of $\sigma$ \cite{Hayashi}. 
From these considerations it would naturally follow that the probability of confusing two quantum 
states $\sigma$ and $\rho$ (after performing $N$ measurements on $\rho$) is (for large $N$):
\begin{equation}
P_N(\rho \rightarrow \sigma) = e^{-N S(\sigma ||\rho)}  \;\; .
\label{main}
\end{equation}
We would like to stress here that classical statistical
reasoning applied to distinguishing quantum states leads to the above formula.
There are, however, other approaches. Some take eq. (\ref{main}) for their 
starting point and then derive the rest of the formalism thenceforth \cite{Petz}.
Others, on the other hand, assume a set of axioms that are necessary to be
satisfied by the quantum analogue of the relative entropy (e.g. it should
reduce to the classical relative entropy if the density operators commute, i.e. if they are ``classical") and then derive eq. (\ref{main}) as a consequence \cite{Donald2}. In any case, as we have argued here, there is a strong reason to
believe that the quantum relative entropy $S(\sigma ||\rho)$ plays the same
role in quantum statistics as the classical relative entropy plays in 
classical statistics. A simple example with a ``quantum coin" will clarify 
this point further \cite{Ekert}. Let us suppose that we have to 
distinguish between a pure, maximally entangled Bell state $|\phi^+\rangle = (|00\rangle + |11\rangle )/\sqrt{2} $ and a mixture $\rho =  (|00\rangle\langle 00| + |11\rangle\langle 11| )/{2}$. Again, we have to decide which state we have by performing $N$ experiments
of our choice on it. In this case we choose to perform projections onto
the state $|\phi^+\rangle = (|00\rangle + |11\rangle )/\sqrt{2}$.
Then if the state $\rho$ is in our possession, we will be successful only $50$ percent of the times (the other $50$ percent of the times we will obtain the orthogonal Bell state $|\phi^-\rangle = (|00\rangle - |11\rangle )/\sqrt{2}$). So, if we perform a single experiment we have a $1/2$ chance of making the wrong inference. If, on the other hand, we have $|\phi^+\rangle$ we will never confuse it with $\rho$ since we are projecting onto the state itself which always gives a positive result. This is in direct analogy with the classical coin example and is, in addition, confirmed by eq. (\ref{main}). In general, however,
the states that we have to distinguish will not be as simple as those above.
Then we would have to find the most optimal measurement to distinguish between
given states in order to reproduce eq. (\ref{main}) from eq. (\ref{lim}).

Now we wish to use the above reasoning to quantify entanglement. Entanglement may be understood as {\bf the distinguishability of a given state from all entirely disentangled ones}. The question is then, in the spirit of the above discussion,
as follows: what is the probability that we confuse a given state with a
disentangled one after performing a total of $N$ measurements? The less the state is entangled the easier it is to confuse it with a disentangled one and vice versa. Thus, the probability to confuse $\sigma$ with a disentangled state, having performed $N$ experiments on $\rho \in {\cal D}$, is of the form

\begin{equation}
e^{-N E(\sigma)} \; ,
\label{pent} 
\end{equation}
where $E(\sigma)$ is the entanglement (obviously if $E=0$ then the state is
indistinguishable from a disentangled one since it is disentangled itself!). 
In comparison with eq. (\ref{main}) we define $E(\sigma)$ to be

\begin{equation}
E(\sigma) := \min_{\rho\in{\cal D}} S(\sigma ||\rho)
\label{ent}
\end{equation}
where ${\cal D}$ is the set of all disentangled states. So for the entanglement of $\sigma$ we use the quantum relative entropy with that disentangled 
$\rho$ which is the most {\bf indistinguishable} from $\sigma$. Obviously, the
greater the entanglement of a state, the smaller is the chance of confusing it with a disentangled state in $N$ measurements. Note that eq. (\ref{ent}) is the same measure as that suggested in our previous letter \cite{Vedral2}. There we showed that the Bures metric, when used instead of $S(\sigma ||\rho)$,
would also be a good measure of entanglement. However, the Bures distance is 
a symmetric quantity and arises from different statistical consideration to those used above (see \cite{Fuchs} for an
overview). Thus, depending on the way we decide to make our measurements, we obtain different ways of comparing the results (i.e. different ``distances" between probability distributions) which, in turn, determine our entanglement
measure (more correctly, the quantity that is to replace $S(\sigma ||\rho
)$ in eq. (\ref{ent})). The convention that we use here assumes performing measurements on $\rho$. We could, of course, envisage making measurements
on $\sigma$, in which case our measure of entanglement would be $E(\sigma) := \min_{\rho\in{\cal D}} S(\rho ||\sigma)$. However, for $\sigma$ being, for example, a maximally entangled Bell state this quantity would be infinite. This agrees
with our statistical interpretation that a Bell state, when measurements are performed on it, could never be confused with a disentangled state and and 
eq. (\ref{pent}) gives probability zero of confusion. But, in order to avoid
dealing with physically undesired infinite amount of entanglement we keep
to the convention given in eq. (\ref{ent}).  

We see that the above treatment does not refer to the number (or indeed dimensionality) of the entangled systems. This is a desired property
as it makes our measure of entanglement universal. However, in order to
perform minimization in eq. (\ref{ent}) we need to be able to define what we
mean by a disentangled state of say $N$ particles. As pointed out in 
\cite{Vedral2} we believe that this can be done inductively.  Namely for
two quantum systems, $A_1$ and $A_2$, we define a disentangled state as one which can be written as a convex sum of disentangled states of $A_1$ and $A_2$ as follows \cite{Horodecki1,Vedral2,Horodecki2}:

\begin{equation}
\rho_{12} = \sum_i p_i \, \rho^{A_1}_i\otimes \rho^{A_2}_i \; ,
\end{equation}
where $\sum_i p_i =1$ and the $p$'s are all positive.
Now, for $N$ entangled systems $A_1, A_2, ... A_N$, the disentangled state is:

\begin{equation}
\rho_{12...N} = \!\!\!\!\!\!\!\!\!\!\!\!\!\! \sum_{\mbox{perm}\{i_1i_2 ... i_N\}} \!\!\!\!\!\!\!\!\!\!\!\!\!\! r_{i_1i_2 ... i_N} \rho^{A_{i_1}A_{i_2} ... A_{i_n}}\otimes 
\rho^{A_{i_{n+1}}A_{i_{n+2}} ... A_{i_{N}}} \; ,
\label{perm} 
\end{equation}
where $\sum_{\mbox{perm}\{i_1i_2 ... i_N\}} r_{i_1i_2 ... i_N} = 1$, all $r$'s are positive and where $\sum_{\mbox{perm}\{i_1i_2 ... i_N\}}$ is a sum over all possible permutations of the set of indices $\{1,2,...,N\}$. 
To clarify this let us see how this looks for $4$ systems:
\begin{eqnarray}
\rho_{1234}  =  & \sum_i & p_i \, \rho^{A_1A_2A_3}_i\otimes \rho^{A_4}_i +  q_i \, \rho^{A_1A_2A_4}_i\otimes \rho^{A_3}_i \nonumber \\
& + & r_i \, \rho^{A_1A_3A_4}_i\otimes \rho^{A_2}_i + s_i \, \rho^{A_2A_3A_4}_i\otimes \rho^{A_1}_i \nonumber \\
& + &  t_i \, \rho^{A_1A_2}_i\otimes \rho^{A_3A_4}_i + u_i \, \rho^{A_1A_3}_i\otimes \rho^{A_2A_4}_i \nonumber \\
& + & v_i \, \rho^{A_1A_4}_i\otimes \rho^{A_2A_3}_i
\label{mess} 
\end{eqnarray}
where, as usual, all the probabilities $p_i,q_i, ..., v_i$ are positive and add up to unity.
The above two equations, at least in principle, define the disentangled states for any number of entangled systems. In practice, unfortunately, this might still not be enough to minimize the relative entropy to obtain the amount of entanglement. So far a good criterion for decomposition into the above form exists for two particles only, when either both are spin $1/2$ or one is spin $1/2$ and the other one is spin $1$ \cite{Horodecki1,Horodecki2} (however, some progress has been made by P. Horodecki \cite{recent}). The above 
definition of a disentangled state is justified by extending the idea that
local actions cannot increase the entanglement between two quantum systems \cite{Bennett1,Vedral1,Vedral2}. In the case of $N$ particles we have $N$
parties (Alice, Bob, Charlie, ... , Wayne) all acting locally on their
systems. The general action that also includes communications can be 
written as \cite{Vedral2}
\begin{equation}
\rho \longrightarrow \!\!\!\!\!\!\! \sum_{i_1,i_2, ... ,I_N} \!\!\!\!\!\!\! A_{i_1}\otimes B_{i_2} \otimes ...
\otimes W_{i_N} \, \rho \, A^{\dagger}_{i_1}\otimes B^{\dagger}_{i_2} \otimes ...\otimes W^{\dagger}_{i_N}
\label{loc}
\end{equation}  
and it can be easily seen that this action does not alter the form of a disentangled state in eqs. (\ref{perm},\ref{mess}). In fact, eq. (\ref{perm}) is the most general state invariant {\bf in form} under the transformation given by eq. (\ref{loc}). We suggest this as a definition of a disentangled state for $N\ge 3$, i.e. it is the most general state invariant in form under local POVM and classical communications. This definition of $N$ particle entanglement means that we say that we do not have $N$ particle entaglement even if subsets of the $N$ particles are individualy entangled. We define it this way so that it answers the question, are all $N$ particles entangled, rather than the question, is there any entanglement at all between the particles. If we wanted to answer the latter question, then clearly the definition of a disentangled $N$ particle state would be one that could be written as
\begin{equation}
   \rho = \sum_{i} p_i \rho_i^A \otimes \rho_i^B \otimes \ldots \otimes \rho_i^W .
\end{equation}

We have in this work derived our previously proposed measure of entanglement from an entirely different perspective. The amount of entanglement is now seen as the quantity that determines ``the least number of measurements that is needed to distinguish a given state from a disentangled one". This therefore strengthens the argument for using eq. (\ref{ent}) as a universal
measure of entanglement. In addition, it opens up the possibility both to understand the meaning of entanglement from a different, more operational, point of view, as well as to measure the amount of entanglement for
more than two quantum systems.  
   
We thank A. Ekert and C. A. Fuchs for discussions and useful comments on this subject.
This work was supported by the European Union, the UK Engineering and Physical Sciences 
Research Council, by a Feodor-Lynen grant of the Alexander 
von Humboldt Foundation, the British Council, the New Zealand Vice Chancellors' Committee and by the Knight Trust.

\end{document}